\documentclass[aps,prl,reprint,showpacs,noshowkeys,floatfix]{revtex4-1}

\usepackage{graphicx}

\usepackage{amsmath,amssymb}

\usepackage[backref=none]{hyperref}
\usepackage[usenames,dvipsnames]{color}
\definecolor{MyDarkBlue}{rgb}{0,0.1,0.7}
\hypersetup{pdfborder={0 0 0},colorlinks,breaklinks=true,
  urlcolor={MyDarkBlue},citecolor={MyDarkBlue},linkcolor={MyDarkBlue}
}

\newcommand{\rmi}{\mathrm{i}}

\newcommand{\hamiltonian}{\hat{\mathcal{H}}}
\newcommand{\field}{\hat{\Psi}}
\newcommand{\cre}[1]{\hat{a}^\dagger_{#1}}
\newcommand{\ann}[1]{\hat{a}_{#1}}
\newcommand{\at}{\tilde{\alpha}}
\newcommand{\nt}{\tilde{N}}
\newcommand*{\diff}[1]{\mathop{}\!\mathrm{d}{#1}\,}

\newcommand{\eref}[1]{(\ref{#1})}
\newcommand{\fref}[1]{Fig.~\ref{#1}}

\begin{document}
\title{Reversible quantum information spreading in many-body systems near criticality}
\newcommand{\RegensburgUniversity}{Institut f\"ur Theoretische Physik, 
Universit\"at Regensburg, D-93040 Regensburg, Germany}
\author{Quirin Hummel}
 \affiliation{\RegensburgUniversity}
\author{Benjamin Geiger}
 \affiliation{\RegensburgUniversity}
\author{Juan Diego Urbina}
\affiliation{\RegensburgUniversity}
\author{Klaus Richter}
\affiliation{\RegensburgUniversity}
\date{\today}

\begin{abstract} 
Quantum chaotic interacting $N$-particle systems are assumed to show fast and irreversible spreading of quantum information on short (Ehrenfest) time scales $\sim\!\log N$. 
Here we show that, near criticality, certain many-body systems exhibit fast initial scrambling, followed subsequently by oscillatory behavior between reentrant localization and delocalization of information in Hilbert space.
We consider both integrable and nonintegrable quantum critical bosonic systems with attractive contact interaction that exhibit locally unstable dynamics in the corresponding many-body phase space of the large-$N$ limit. 
Semiclassical quantization of the latter accounts for many-body correlations in excellent agreement with simulations.
Most notably, it predicts an asymptotically constant local level spacing $\hbar/\tau$, again given by $\tau\! \sim\! \log N$.
This unique timescale governs the long-time behavior of out-of-time-order correlators that feature quasi-periodic recurrences indicating reversibility.
\end{abstract}


\maketitle

The dynamics of quantum information in complex many-body (MB) systems presently attracts a lot of attention \cite{Altman2018,Swingle2018} ranging from atomic and condensed quantum matter to high energy physics. 
The evolution of an (excited) quantum MB system towards a state of thermal equilibrium usually goes along with the scrambling of quantum correlations, encoded in the initial state, across the system's many degrees of freedom.
Such dynamics requires an improved understanding of MB quantum chaos and the link with thermalization \cite{Srednicki1994,Rigol2008,Eisert2015,Kaufman2016} and its suppression \cite{Nandkishore2015,Schreiber2015,Altman2018}.

Echo protocols, measuring how a perturbation affects successive forward and backward propagations in time, sensitively probe the stability of complex quantum dynamics.
Here, out-of-time-order correlators (OTOCs) \cite{Larkin1969,Maldacena2016,Maldacena2016b}
\begin{equation} 
 \label{eq:OTOC-general}
  C(t) = \langle [\hat{W}(t), \hat{V}]^\dagger [\hat{W}(t), \hat{V}] \rangle
\end{equation}
play a central role, with first experimental implementations \cite{Gaerttner2017,Li2017,Wei2018}, allowing to distinguish various classes of MB systems by their operator growth.
On the one side there are slow scramblers, such as systems in the MB localized phase exhibiting logarithmically slow operator spreading \cite{Chen2016,Huang2017,Fan2017,Swingle2017} or, e.g., Luttinger liquids \cite{Dora2017} showing only quadratic increase.
On the other side, an exponentially fast initial growth of OTOCs is commonly viewed as a quantum signature of MB chaos. 
Examples comprise systems with holographic duals to black holes \cite{Maldacena2016,Cotler2017}, the SYK-model \cite{Kitaev2015,Maldacena2016b,Polchinski2016,Patel2017}, and condensed matter systems close to a quantum phase transition (QPT) \cite{Shen2017,Heyl2018,Alavirad2019,Chavez-Carlos2019} or exhibiting chaos in the classical limit of large particle number $N$.
In such large-$N$ systems, the exponential growth rate for OTOCs is given by the Lyapunov exponent of their classical counterpart \cite{Maldacena2016,Swingle2016,Rozenbaum2017,Bohrdt2017,Scaffidi2017,Rammensee2018,Garcia-Mata2018,Chavez-Carlos2019,Jalabert2018} and prevails up to the Ehrenfest $\log N$ time where MB quantum interference sets in \cite{Rammensee2018,Tomsovic2018}.
Subsequent OTOC time evolution towards an ergodic limit is then often governed by slow classical modes \cite{Sondhi2018}.

Here we show that exponentially fast scrambling need not necessarily lead to quantum information loss: 
There exist systems exhibiting initial growth of complexity without relaxation, i.e.,  after a quench to an interacting system close to criticality the OTOCs do not show monotonous saturation; instead the correlations imprinted initially can be periodically retrieved.

Quantum critical large-$N$ systems are particularly sui\-ted for considering the inter-relation between spreading of correlations, quantified through OTOCs, and corresponding nonlinear  classical mean-field (MF) dynamics.
There, critical phenomena are often viewed as quantum manifestations of structural changes in classical phase space, associated with unstable MF motion close to separatrices. While corresponding studies \cite{Emary2003,Caprio2008,Bastidas2014,Stransky2014,Bastarrachea-Magnani2016,Rubeni2017,Pappalardi2018} commonly invoke a classical MF analysis, we will show that MB semiclassical quantization beyond MF allows for a precise characterization of the locally unstable quantum dynamics.
In the language of renormalization group analysis we therefore expect our results to be valid for any dimension within lower and upper critical dimension as long as a MF (classical) limit exists \footnote{Following \cite{Chaikin1995} criticality occurs above the lower, while quantum fluctuations become subdominant above the upper critical dimension.}. %
\nocite{Chaikin1995}

While in generic quantum chaotic systems the Ehrenfest time $\sim\log N$ is distinctly shorter than the Heisenberg time (associated with the inverse mean level spacing) we will show that these two scales indeed coincide for certain quantum critical systems where an adiabatic separation allows for an effective 1D description.
Quantization of their locally hyperbolic MB dynamics implies two inter-related features: 
Even though the dynamics may be separable, OTOCs still grow exponentially with a rate given by the local MF instability exponent $\lambda_s$ up to times $(1/\lambda_s)\log N$.
Second, the inverse mean level spacing in the relevant spectral region also scales as $\log N$.
Hence, the quantum critical dynamics is governed by $\log N$ as the sole time scale. 

Remarkably, this level spacing turns out to be asymptotically constant, approaching a harmonic oscillator spectrum, although the underlying hyperbolic dynamics is unstable and rather corresponds to an inverted oscillator
\footnote{%
See~\cite{Molina-Vilaplana2013} for similar results for the $xp$ model in $\mathrm{AdS}_2$.%
}.
This equidistant level spacing implies strong, periodic quantum recurrences on short $\log N$-scales that dominate OTOCs and hence reflect {\em unscrambling} of information in quantum critical MB systems. 
After showing this behavior in a prototypical integrable model we consider a nonintegrable extension and confirm the robustness of this feature, indicating that it is not linked to integrability but is characteristic for QPTs driven dominantly by a single degree of freedom.
On the contrary, in generic chaotic MB systems randomlike evolution is expected for enormously long (Heisenberg) times beyond which the spectral discreteness eventually enforces recurrences
\footnote{%
Such recurrences are exceptional and exist only at particular wavelengths \cite{Tomsovic1997}.%
}.

{\em Quantum critical atomic Bose gas}.---%
As a generic example of critical behavior we consider the 1D attractive Bose gas with periodic boundary conditions (attractive Lieb-Liniger model) \cite{Lieb1963,Kanamoto2003,Sykes2007}.
While its Hamiltonian
\begin{equation} \label{eq:hamiltonian}
 \hamiltonian=\int_0^{2\pi}\diff{\theta} \field^\dagger(\theta)(-\partial^2_\theta)\field(\theta)
	-\frac{\pi \at}{2} \big[\field^\dagger(\theta)\big]^2\big[\field(\theta)\big]^2,
\end{equation}
with bosonic field operators $\field$ and $\field^\dagger$, describes quasi-1D ultracold atomic gases with interactions parametrized by $\at$ \cite{Strecker2002,Khaykovich2002,Chin2010}, its MF dynamics is governed by the Gross-Pitaevskii equation.
It exhibits a QPT at a critical coupling $\at N=1$ \cite{Kanamoto2003,Kanamoto2005,Kanamoto2006} where the homogeneous condensate starts forming a bright soliton.
Although for finite $N$ eigenvalues of the quantum integrable Hamiltonian (\ref{eq:hamiltonian}) can be, in principle, found through Bethe ansatz \cite{Sakmann2005,Sykes2007}, this does not allow for systematically treating the $N\to\infty$ limit, except for special states \cite{Flassig2016,Piroli2016}.
Instead we first truncate \cite{SMmodel}
%
%
%
\nocite{SMmodel,SMclassical,SMunivSepQuant,SMsepQuantModel,SMequidistance,SMlogtimeEntropy,SMfiveMode}%
%
%
%
$\field(\theta)$ to the lowest three momentum modes,
\begin{equation} \label{eq:3mod}
 \field(\theta)=(\ann{0}+\ann{-1}e^{-\rmi\theta}+\ann{1}e^{\rmi\theta})/ \sqrt{2\pi} \, , 
\end{equation}
as commonly done for exact diagonalization \cite{Kanamoto2005,Dvali2013,Dvali2015}---a good approximation for $\at N<1$ that also contains all the physics relevant for understanding the QPT and its precursors for $\at N\geq1$ \cite{Kanamoto2003,Kanamoto2006,Sykes2007,Dvali2013}.
The generalization to the non-integrable 5-mode model will be discussed later.
The model (\ref{eq:hamiltonian},\ref{eq:3mod}) near the QPT mimics black holes as graviton condensates \cite{Dvali2013,Dvali2014,Dvali2015}, can be essentially realized using ultracold spin-1 atoms \cite{Gerving2012} and has attracted considerable attention~\cite{Arwas2015,Pruefer2018,Garcia-March2018} for time crystals \cite{Kosior2018}. 

\begin{figure}
	\includegraphics[width=\linewidth]{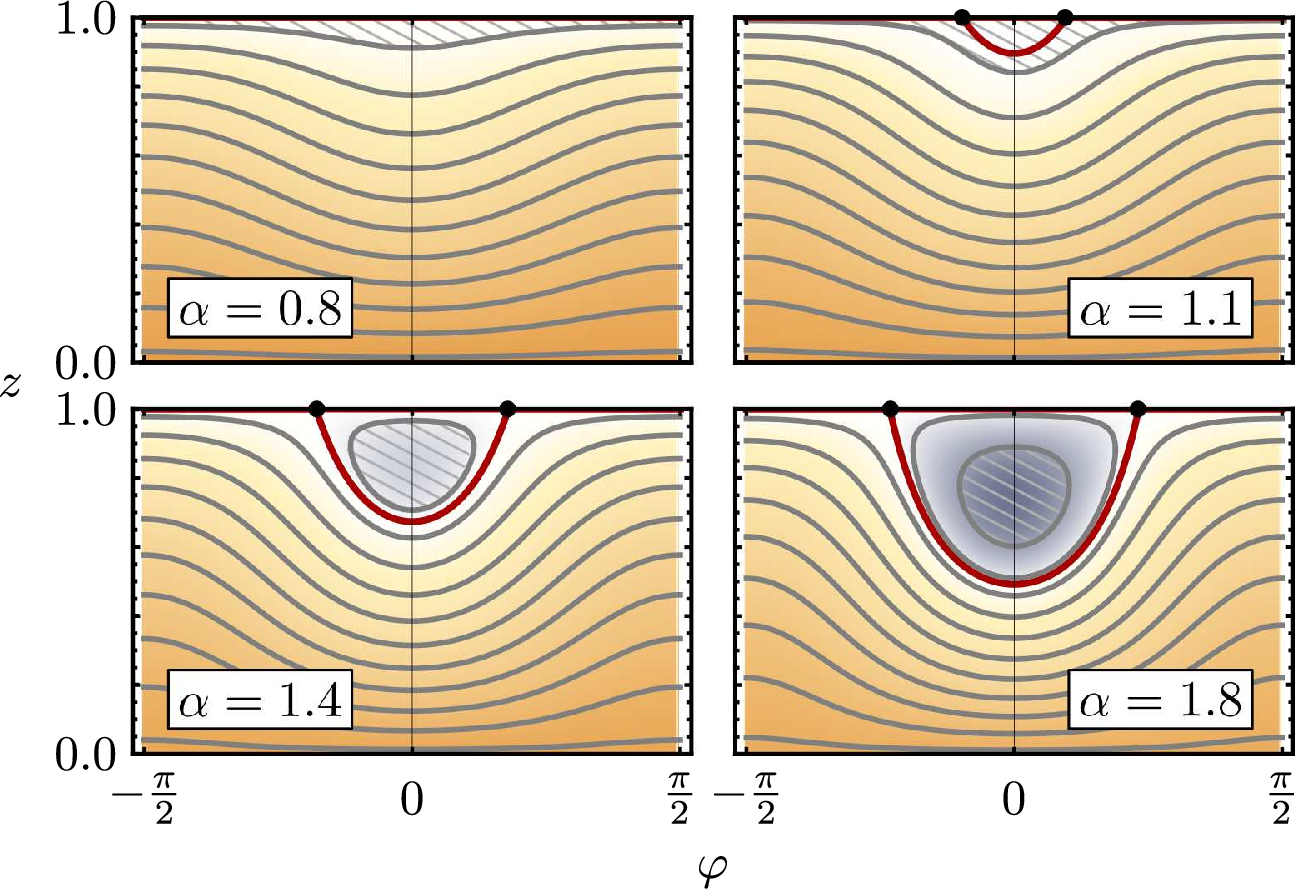}
   \caption{Phase space portrait of energy $\omega(z,\varphi)$, Eq.~(\ref{eq:omega}), of the classical dynamics for the three-mode model of attractive bosons in 1D. $z$ denotes the relative occupation of the noninteracting ground state and $\varphi$ the conjugate angle, for different values of interaction $\alpha$. 
		For $\alpha>1$ a global energy minimum (cross) and a separatrix (red) connecting two hyperbolic fixed points (dots at $z=1$) appear.
		Gray lines represent the orbits (tori) that fulfill the quantization condition (\ref{eq:QuantCond}) for $N\!=\!20$.
		Note that the quantized orbits change their character sequentially.
	}
	\label{fig:quantized_orbits}
\end{figure}

{\em Classical MF limit}.---%
Besides the energy also particle number and total (angular) momentum are conserved:
\begin{equation} 
\label{eq:NandK}
\hat{N}=\sum_{k}\cre{k}\ann{k} \,,  \quad 
\hat{K}= \sum_k  k \, \cre{k}\ann{k}  \, .
\end{equation}
Hence, the truncation to three modes, in contrast to five or more modes \cite{Herbst1989}, renders the system integrable in that its large-$N$ MF limit, formally representing a classical limit, is integrable.
This allows for devising a MB version of semiclassical torus quantization \cite{Tabor1989,OzorioDeAlmeida1990} to analytically find the spectrum and wave functions being asymptotically exact for $N\! \to \! \infty$.
To this end we write the operators in symmetric order and replace $\ann{k}\mapsto \sqrt{n_k}e^{\rmi\vartheta_k}$ for $k\!=\! -1,0,1$, where $(n_k,\vartheta_k)$ are continuous classical conjugate variables. 
Using
\begin{equation} \label{eq:NandK-class}
	\nt\!\equiv n_{-1}+ n_0+n_1 \,, \quad  K\equiv n_1-\!n_{-1} 
\end{equation}
and considering $K=0$, the classical energy per particle is, defining $\alpha=\at \nt$, \cite{SMclassical}
\begin{equation} \label{eq:EpP}
	\frac{E}{\nt}=\omega(z,\varphi)+
        \alpha\left(-\frac{1}{4}+ \frac{3}{2\nt}-\frac{9}{8\nt^2} \right) -\frac{1}{\nt} \,,
\end{equation}
where the classical dynamics is completely determined by
\begin{equation} \label{eq:omega}
	\omega(z,\varphi)=(1-z)\left[1-\alpha\left[(1-z)/8 +z\cos^2\varphi\right]\right]
\end{equation}
with phase space coordinates
\begin{equation} \label{eq:scale}
	z=\frac{n_0}{\nt}, \quad \varphi=\vartheta_0-\frac{1}{2} (\vartheta_1+\vartheta_{-1}) ,\quad
\{z,\varphi\}=\frac{1}{\nt} \, .
\end{equation}
Note that $1/\nt$ only enters as an effective quantum of action in the Poisson bracket \cite{Engl2014}.

The Hamiltonian $\omega(z,\varphi)$ involves different types of classical trajectories following lines of constant energy in phase space with periodicity $\varphi\!\mapsto\!\varphi+\pi$, see Fig.~\ref{fig:quantized_orbits}. 
For $\alpha<1$ all trajectories are deformed horizontal lines (rotations). 
For $\alpha>1$ an island centered around a new minimum energy fixed point emerges with orbits vibrating in $\varphi$, similar as for the pendulum.
This goes along with the formation of a separatrix at $E\!=\!E_\mathrm{sep}$ ($\omega\!=\!0$) 
associated with two hyperbolic fixed points at $z\!=\!1$  and
characterized by (in)stability exponents 
\begin{equation} 
\label{eq:lambda-s}
\lambda_\mathrm{s}^{(1,2)} = 2\sqrt{\alpha-1} \equiv \lambda_\mathrm{s} \, .
\end{equation}

\begin{figure}
	\includegraphics[width=\linewidth]{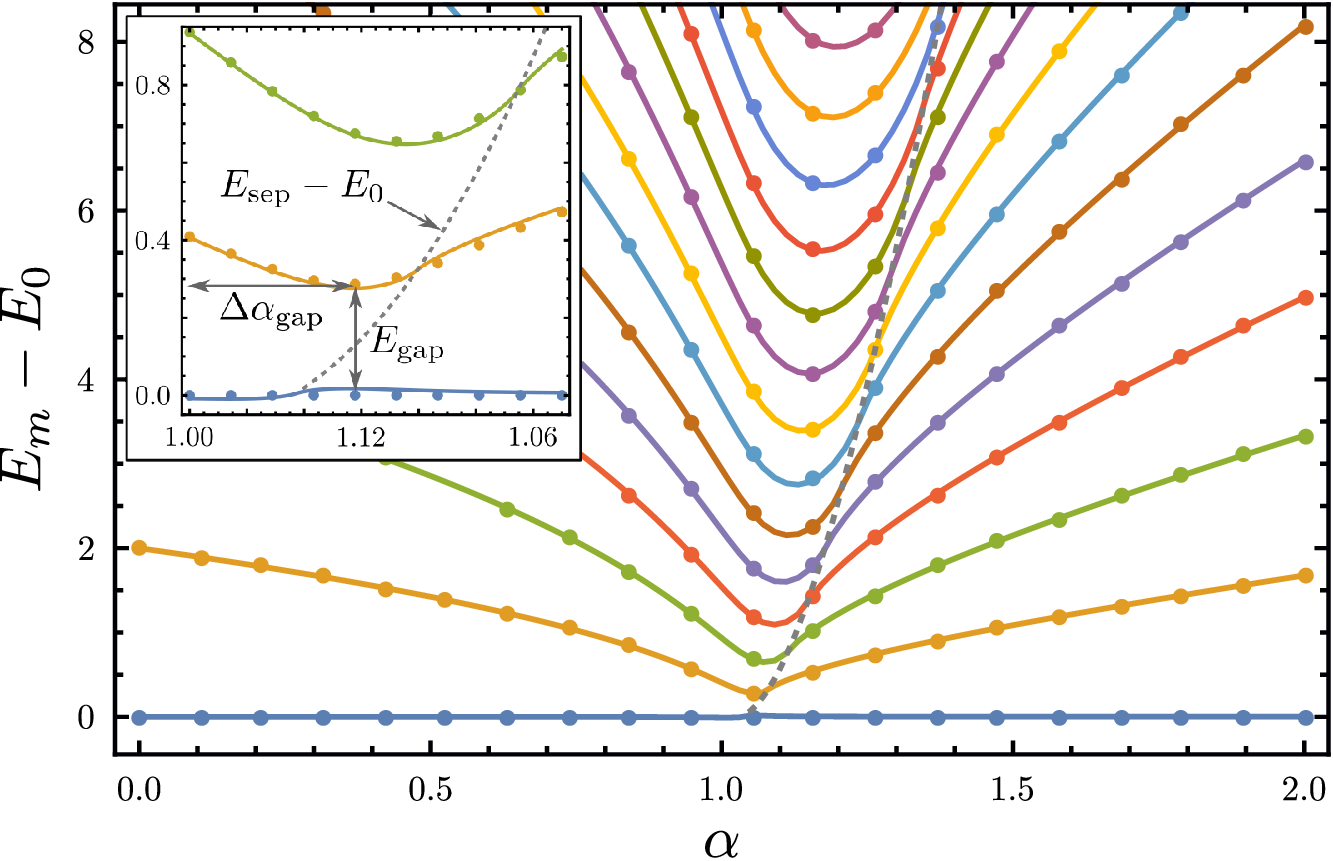}
	\caption{Excitation spectrum $E_m \! -\! E_0$ for $N\! = \! 300$ and $0 \le \alpha \le 2$ within 3-mode approximation, Eq.~(\ref{eq:3mod}). 
         Symbols and lines denote numerical and semiclassical results, respectively. The gray dotted line indicates inflection points, the finite size precursor of an excited state QPT, where quantized 
         orbits cross the separatrix (in Fig.~\ref{fig:quantized_orbits}).
         Inset: closeup around $\alpha\!=\! 1$.}
	\label{fig:ExSpec}
\end{figure}

{\em Semiclassical quantization}.---%
To study genuine quantum effects we go beyond the classical MF picture using semiclassical torus quantization. 
While related WKB approaches, devised for one-dimensional systems, were successfully used in two-site models \cite{Albiez2005,Gati2006,Graefe2007}, we adapt a multi-dimensional generalization.
In the MB context it yields the quantization rules
\begin{eqnarray} \label{eq:QuantCond}
		& &  \frac{1}{2\pi} \oint  
\diff{\varphi}[1-z(\omega,\varphi)] = \frac{m+\frac{1}{2}}{\nt} \,, \\
 		& & m =  0,1,\dots \lfloor N/2 \rfloor ,  \quad  \nt=N+3/2 , \quad  N=0,1,\dots \,. \nonumber
\end{eqnarray}
Equations (\ref{eq:QuantCond}) effectively quantize the phase space areas bounded by the lines $\omega=\mathrm{const.}$ (shaded areas in Fig.~\ref{fig:quantized_orbits} for $m=0$), giving rise to energies $\omega_m \!=\! E_m/ \nt$, in perfect agreement with results from exact diagonalization, see Fig. \ref{fig:ExSpec}.

{\em Quantum phase transition}.---%
The QPT for $N\!\to\!\infty$ is associated with the ground state corresponding to the quantized orbit enclosing the phase space area $1/(2\nt)\to0$ that is always vibrational for $\alpha>1$ if $\nt$ is large enough. 
Its energy scales as $\omega_\mathrm{min}\sim-(\alpha-1)^2$, in contrast to $\omega_\mathrm{min}=0$ for $\alpha<1$ where the quantized orbit approaches $z=1$, leading to the nonanalytic dependence on $\alpha$ of the MF ground state at $\alpha\!=\!1$. 
Precursors of such nonanalyticity for finite $\nt$ appear for every quantized orbit changing from rotation to vibration upon tuning $\alpha$.
This is reflected in the sequence of avoided crossings in Fig.~\ref{fig:ExSpec} building up an excited state QPT when $\nt \to \infty$ \cite{Caprio2008,Bastarrachea-Magnani2016}.
Remarkably, Eq.~(\ref{eq:QuantCond}) even allows us \cite{Hummel2017thesis} to analytically obtain the scaling laws
\footnote{%
We are not aware of renormalization group calculations of critical exponents in this model.
} 
governing the approach of $\Delta\alpha_\mathrm{gap}\sim N^{-2/3}$ and $E_\mathrm{gap}\sim N^{-1/3}$ (see inset of Fig.~\ref{fig:ExSpec}) to their MF values, in perfect agreement with numerical and heuristic observations \cite{Kanamoto2003,Kanamoto2005,Dvali2015}.
 
\begin{figure} 
	\includegraphics[width=\linewidth]{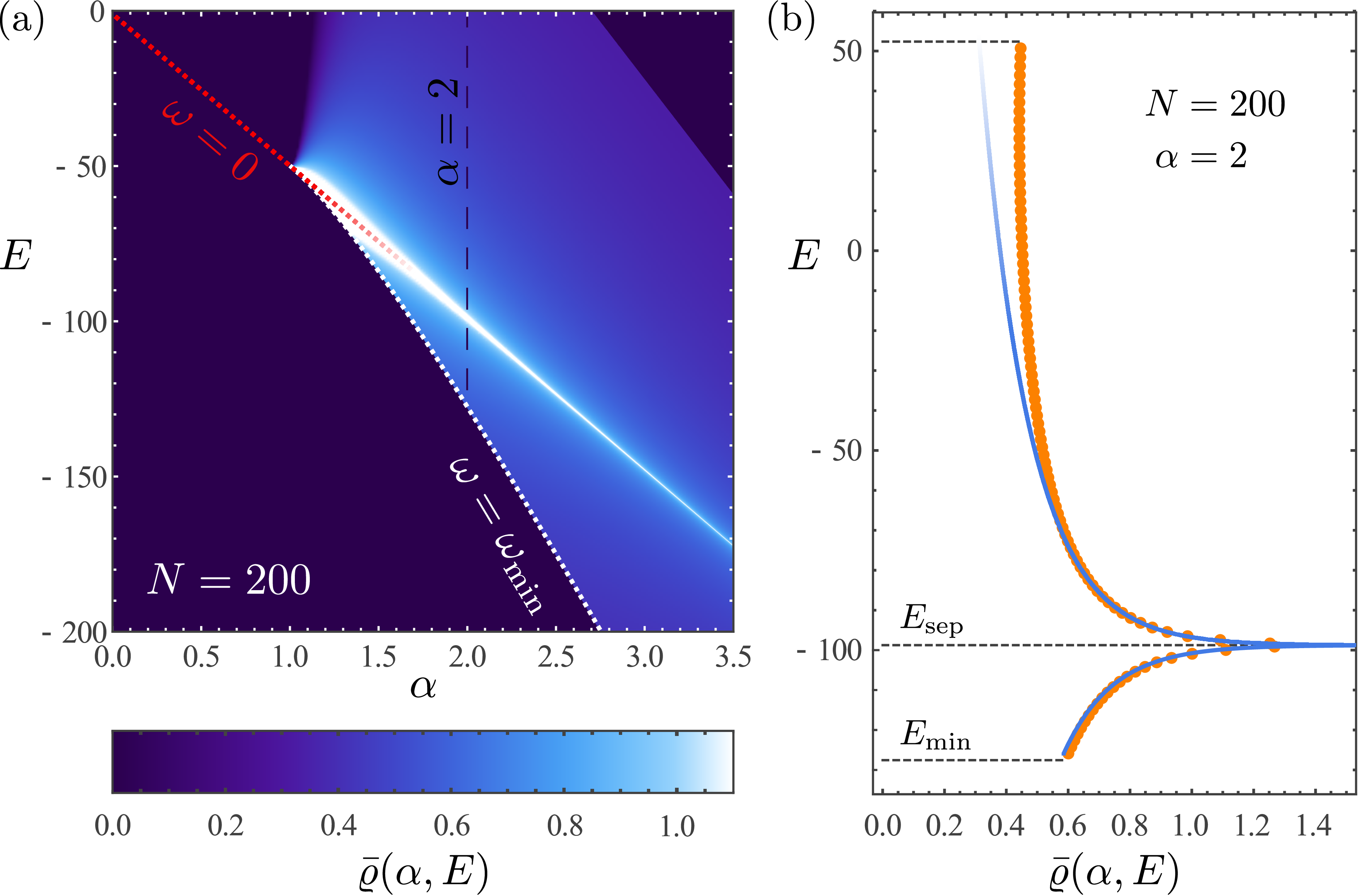}
	{\caption{\label{fig:DOSasym}
		(a) Asymptotic $N\gg 1$ density of states $\bar{\rho}(\alpha,E)$, Eq.~\eref{eq:DOSasym}, showing level bunching (bright straight line) around the precursor of the excited state QPT at $E\!=\! E_{\rm sep}$ in the supercritical regime $\alpha \! > \!  1$.\
		(b) Slice of  $\bar{\rho}(\alpha,E)$ at $\alpha \!=\! 2$ compared with numerically calculated inverse level gaps (orange symbols) exhibiting a characteristic logarithmic divergence.
	}}
\end{figure}

Hence, MB semiclassical quantization goes beyond the Bogoliubov picture of \cite{Kanamoto2003,Kanamoto2006} where the excitation spectrum collapses to zero at the MF critical coupling $\alpha \!=\! 1$. 
Instead, for finite $\nt$ we find huge accumulations of levels around the separatrix.
This precursor of an excited state QPT leads to characteristic features in the spectra and to the emergence of a {\it local log time scale}.
An asymptotic $N\gg1$ analysis \cite{SMsepQuantModel} of \eref{eq:QuantCond} that generally holds close to a separatrix \cite{SMunivSepQuant} yields the average density of states
\begin{equation} \label{eq:DOSasym}
	\bar{\varrho}(E) \!=\!  \frac{-1}{2\pi \lambda} \log \!\left( \frac{\left|E \!-\! E_{\rm sep}\right|}{\tilde{N}}
\right) + \mathcal{O}(1)
\end{equation}
with a characteristic logarithmic divergence at $E=E_{\rm sep}$ (see \fref{fig:DOSasym}).
Here $\lambda \!=\! \lambda_\mathrm{s} / 2  \!=\! \sqrt{\alpha -1}$, Eq.~(\ref{eq:lambda-s}), 
and the $\mathcal{O}(1)$ finite size correction involves a system specific timescale related to the traversal along the separatrix \cite{SMunivSepQuant,SMsepQuantModel}.
\begin{figure}
	\includegraphics[width=\linewidth]{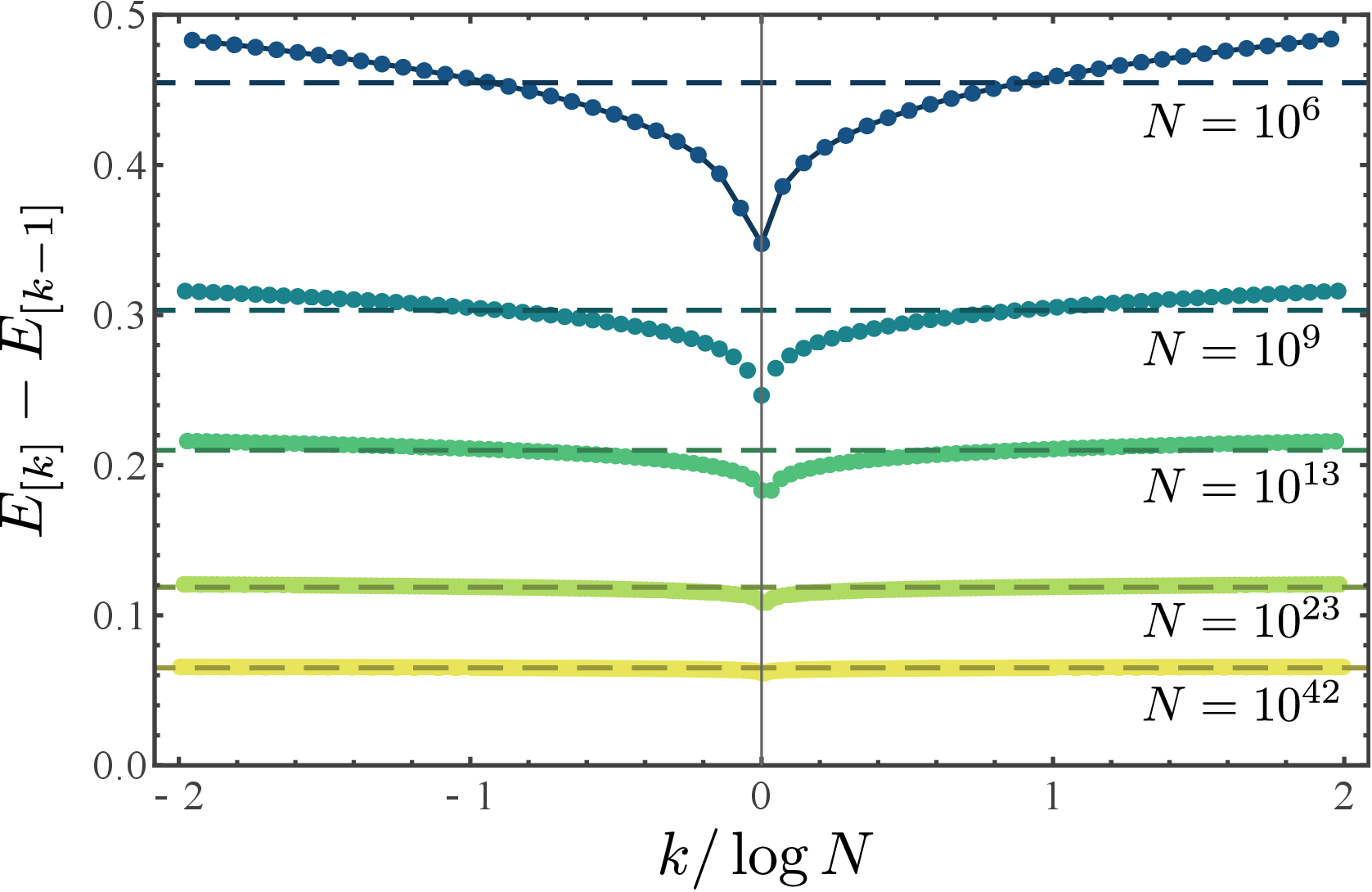}
	{\caption{\label{fig:equidistant}
		Asymptotic level spacings $E_{[k]} - E_{[k-1]}$ (symbols) obtained from \eref{eq:DOSasym} \cite{SMsepQuantModel}, where $E_{[k=0]}$ refers to the level closest to the separatrix energy (excited state QPT), approaching the characteristic constant spacing $\Delta E$, Eq.~\eref{eq:tau}, (dashed) for $\alpha=2$ and $N=10^6,10^9,10^{13},10^{23},10^{42}$ (the latter with relevance for black hole physics) from top/blue to bottom/yellow within windows with $\sim\! \log N$ levels.
		Energies obtained from full semiclassical quantization, \eref{eq:QuantCond}, (solid line) for $N=10^6$ confirm the validity of Eq.~\eref{eq:DOSasym}.
	}}
\end{figure}

{\em Asymptotically constant level spacing and log time}.---%
Most notably, evaluating (\ref{eq:QuantCond}) close to $E_{\rm sep}$ one finds \cite{SMequidistance} that a set of levels, growing in number logarithmically with $N$, becomes {\it asymptotically equidistant} with level spacing $\Delta E$, see \fref{fig:equidistant}.
The associated time scale \cite{SMlogtimeEntropy}
\begin{equation} \label{eq:tau}
	\tau = \frac{2\pi}{\Delta E} = \frac{1}{\lambda} \log \tilde{N} + \mathcal{O}(1)
\end{equation}
is the Heisenberg time corresponding to the local spectral gap $\Delta E$, but exhibits a striking similarity with the Ehrenfest time $\tau_{\rm E} \!=\! (1 / \lambda_{\rm L}) \log \hbar_{\rm eff}^{-1}$ with $\hbar_{\rm eff} \! = \hbar$ and $1/\nt$ in chaotic single-particle \cite{Berman1978}, respectively, MB systems \cite{Rammensee2018} with Lyapunov exponent $\lambda_{\rm L}$.
This justifies to relate \eref{eq:tau} to a \textit{local Ehrenfest time} associated with the dynamical instability characteristic of critical behavior.
Due to the universality of \eref{eq:DOSasym} and \eref{eq:tau}, supported by the classical renormalization group analysis of \cite{Zaslavsky2005}, this turns out to be a generic behavior close to hyperbolic fixed points.
The crossover of the Heisenberg time from usually algebraic in $N$ to log time behavior is not shared by generic chaotic systems.

{\em Out-of-time-order correlator}.---%
We address the drastic consequences of this transition for the evolution of quantum MB correlations: 
We quantify the spreading of information through the OTOC
, Eq.~(\ref{eq:OTOC-general}),
\begin{equation}
	C(t)=-N^{-4}\langle\psi|[\hat{n}_0(0),\hat{n}_0(t)]^2|\psi\rangle
	\label{eq:OTOC}
\end{equation}
for operators $\hat{n}_0(t)/N$.
In chaotic systems quasi-classical arguments \cite{Larkin1969,Maldacena2016,Swingle2016} confirmed by MB semiclassical theory \cite{Rammensee2018} predict a short-time behavior $C(t)\sim \hbar^2e^{2\lambda_\mathrm{L}t}$ passing into a saturation regime at $\tau_E$.
Although the system (\ref{eq:3mod}) is integrable, we can use similar arguments to predict that in the quantum critical regime the dynamical instability close to the hyperbolic fixed points also causes such an exponential behavior,
\begin{equation}
	C(t)\sim (\hbar_\mathrm{eff}^2 /\nt^2)\, e^{2\lambda_\mathrm{s}t},
	\label{eq:OTOCscaling}
\end{equation}
but with a rate given by the (in)stability exponents $\lambda_\mathrm{s}$, Eq.~(\ref{eq:lambda-s}) (see \cite{Pappalardi2018} for a related result for chains of large spins).

\begin{figure}
	\includegraphics[width=\linewidth]{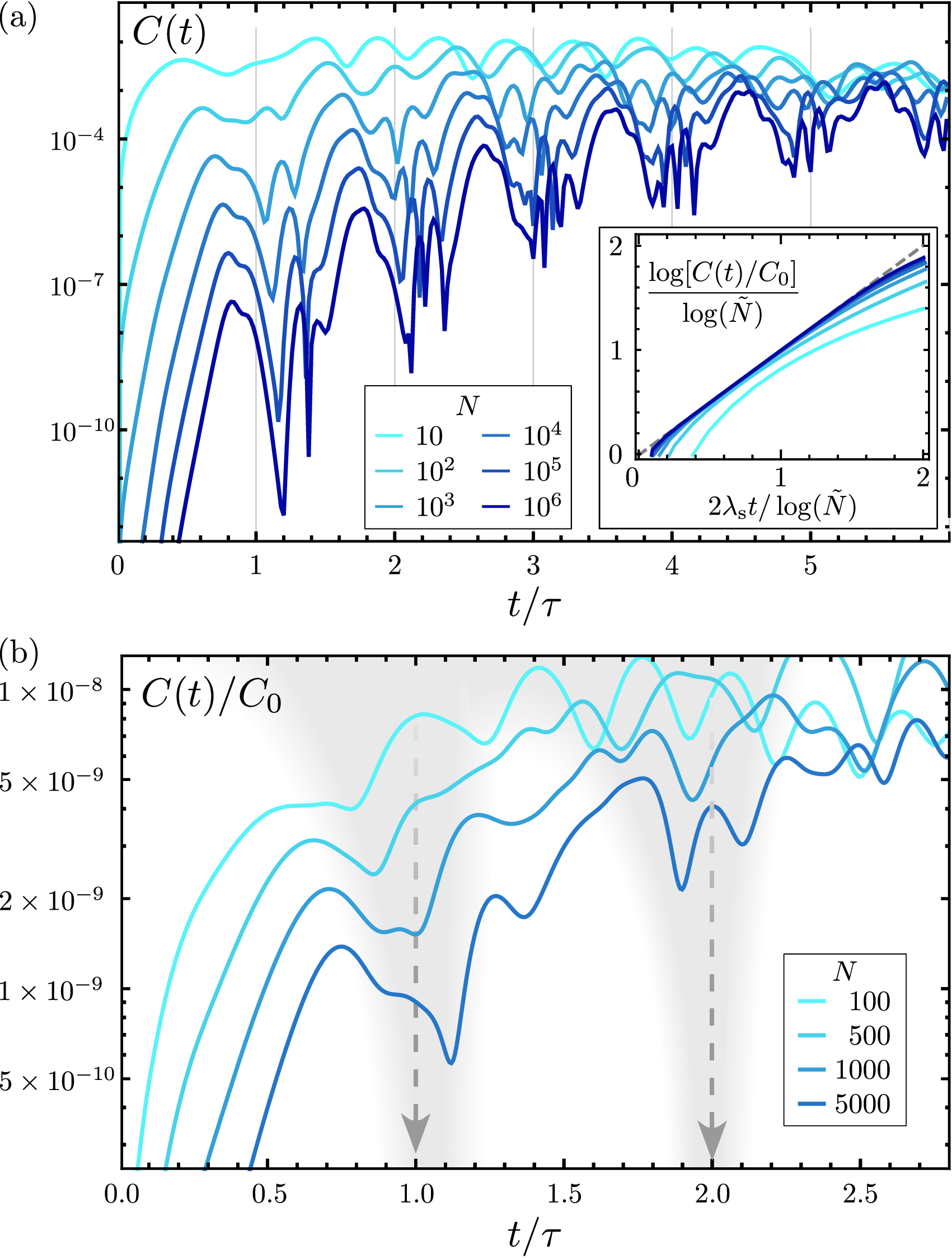}
	\caption{
		Oscillatory time evolution of the OTOC \eref{eq:OTOC} after a quench of the noninteracting condensate across the mean-field critical point, reflecting scrambling and unscrambling (i.e., indicating reversibility) close to criticality for various particle numbers $N$.
		(a) $C(t)$ exhibits distinct, approximately $\tau$-periodic oscillations where $\tau$, Eq.~\eref{eq:tau}, is the local Ehrenfest time [3-mode model (\ref{eq:3mod}) with $\alpha=2$].
		Inset: Initial growth of $C(t)$ approaching the exponent $2\lambda_\mathrm{s}t$ (dashed line) with increasing $N$, thereby confirming Eq.~(\ref{eq:OTOCscaling}).
		(b) Tendency towards periodicity (suggested in gray) for the non-integrable extended 5-mode model ($|k| \leq 2$, see \cite{SMfiveMode}) for $\alpha = 1.05$, with a period of the form \eref{eq:tau} with $\lambda \approx 0.21$, consistently found in the level spacings and the classical instability.
		The prefactors $C_0$ facilitate the comparison.%
	}
	\label{fig:OTOC}
\end{figure}

Within the present MB model 
we have numerical access to huge particle numbers and hence can thoroughly check the commonly assumed 
exponential growth of OTOCs in the truly semiclassical large-$N$ limit, as well as associated log-time 
effects.
In Fig.~\ref{fig:OTOC} we present numerical results for $C(t)$ computed from (\ref{eq:OTOC}) after
imposing an interaction quench to the non-interacting ground state 
$|\psi \rangle = 1/ \sqrt{N!} (\hat{a}_0^\dagger)^N | 0 \rangle$.
The inset displays the short time behavior of $\log C(t)$ up to the Ehrenfest time scale $ \tau_E = (1/\lambda_\mathrm{s}) \log \nt$
for $N\!=\!10^2$ to $10^6$, showing convergence to the slope $2\lambda_\mathrm{s}t$ (dashed line) predicted in (\ref{eq:OTOCscaling}).

{\em Unscrambling of correlations}.---%
Figure \fref{fig:OTOC}(a) shows that instead of monotonously approaching a plateau for $t>\tau_\mathrm{E}$, as in generic chaotic MB systems,
$C(t)$ exhibits distinct oscillations with a period given by the log time $\tau$, \eref{eq:tau}, that includes finite-size corrections \cite{SMlogtimeEntropy} and hence slightly differs from $\tau_E$.
Since $C(t)$ is a measure of information spreading, these oscillations reflect reversibility of quantum information flux in Hilbert space as a result of genuine MB interference.
This is supplemented by corresponding oscillations in the evolution of entanglement encoded in the one-body entropy \cite{SMlogtimeEntropy}.
Close to criticality, $C(t)$ is dominated by an increasing (with $\log N$) number of states close to $E_{\rm sep}$ \cite{SMlogtimeEntropy} where the spectrum gets asymptotically equidistant (Fig.~\ref{fig:equidistant}).
This induces revivals (getting more and more pronounced with increasing $N$) associated with a unique time scale, the log time $\tau$, (\ref{eq:tau}), taking the role of a Heisenberg time close to criticality.
To clearcut show this asymptotic periodicity, here deduced from MB separatrix quantization, requires large-$N$ regimes that, to the best of our knowledge, are not accessible with present numerical methods for chaotic MB systems.

To assess the latter and to explore the generality of our findings we further relax the truncation \eref{eq:3mod} to five modes ($|k|\leq2$), implying unstable non-integrable dynamics.
Within the bounds of numerical tractability ($N=\mathcal{O}(10^4)$) we verify that this quantum critical MB system, despite being non-integrable, exhibits the same tendency towards periodic revivals, see Fig.~\ref{fig:OTOC}(b).
Via adiabatic separation we can again attribute this behavior to locally hyperbolic MF dynamics of a single dominant degree of freedom \cite{SMfiveMode}.
Moreover, we find the same interrelation between the MF instability $\lambda_{\mathrm{s}}$, the period $\tau$, and the local level spacing as in the integrable three-mode model.%

In conclusion, by means of many-body semiclassical quantization we could explain the exponentially fast scrambling and buildup of correlations in  quantum systems that are critical, but not globally chaotic.
We uncovered a generic me\-cha\-nism for fast dynamics near a quantum critical point.
Moreover, for large $N$ we demonstrated the emergence of nearly equidistant spectra giving rise to recurrences in OTOCs
on $\log N$-time scales,
resembling features of time crystals~\cite{Kosior2018}.
Their observation should be in experimental reach since, e.g., recurrences based on stable dynamics have already been observed in a system with thousands of atoms~\cite{Rauer2018}.
Our analysis of the non-integrable 5-mode model shows that such memory effects do not require integrability and indicates their existence in larger classes of critical systems with a dynamical decoupling of a dominant unstable mode from other collective degrees of freedom.
Moreover, our results shed light on generic mechanisms governing the dynamics at (excited-state) quantum phase transitions beyond mean-field.

We acknowledge funding through the Studien\-stiftung des Deutschen Volkes (BG) and the Deutsche Forschungsgemeinschaft through project Ri681/14-1.

\nocite{Molina-Vilaplana2013,Tomsovic1997}

\nocite{Castin2004,Lieb1963,Olshanii1998,Nishida2018,Pricoupenko2018,Guijarro2018,Kanamoto2005,Kanamoto2006,Sykes2007,Gardiner2004,Hummel2017thesis,Pitaevskii2016,Bogoliubov1947,deGennes1966,Fetter1972,Kanamoto2003,Dvali2013,Stransky2014,Caprio2008,Corless1996,Dvali2013,Mathew2017}

\bibliography{bibliography}
\end{document}